\documentclass[12pt,twoside]{article}
\usepackage[mathscr]{eucal}
\usepackage{amsmath,amsfonts,amssymb,amsthm,mathabx}
\usepackage{times}
\usepackage{pdfsync}
\usepackage{cite}
\usepackage{url}
\usepackage{hyperref}
\usepackage{tensor}
\usepackage{color}
\usepackage{caption}

\advance\textheight\topskip
\numberwithin{equation}{section} \makeatletter
\@addtoreset{equation}{section}

\newcommand{\D}{\text{d}}

\begin{document}

\title{Modular invariance in finite temperature Casimir effect}

\author{Francesco Alessio, Glenn Barnich}


\def\mytitle{Modular invariance in finite temperature Casimir effect}

\pagestyle{myheadings} \markboth{\textsc{\small F.~Alessio,
    G.~Barnich}}{\textsc{\small Modular invariance in Casimir effect}}

\addtolength{\headsep}{4pt}

\begin{centering}

  \vspace{1cm}

  \textbf{\Large{\mytitle}}

  \vspace{1.5cm}

  {\large Francesco Alessio}

  \vspace{.5cm}

\begin{minipage}{.9\textwidth}\small \it \begin{center} Dipartimento di Fisica ``E. Pancini'' and INFN\\
    Universit\`a degli studi di Napoli ``Federico II'', I-80125
    Napoli, Italy
\end{center}
\end{minipage}

\vspace{1cm}

{\large Glenn Barnich}

\vspace{.5cm}

\begin{minipage}{.9\textwidth}\small \it \begin{center} Physique
    Th\'eorique et Math\'ematique \\ Universit\'e libre de Bruxelles
    and International Solvay Institutes \\ Campus Plaine C.P. 231,
    B-1050 Bruxelles, Belgium
\end{center}
\end{minipage}

\end{centering}

\vspace{1cm}
  
\begin{center}
  \begin{minipage}{.9\textwidth} \textsc{Abstract}. The temperature inversion
    symmetry of the partition function of the electromagnetic field in the
    set-up of the Casimir effect is extended to full modular transformations by
    turning on a purely imaginary chemical potential for adapted spin angular
    momentum. The extended partition function is expressed in terms of a real
    analytic Eisenstein series. These results become transparent after explicitly
    showing equivalence of the partition functions for Maxwell's theory between
    perfectly conducting parallel plates and for a massless scalar with periodic
    boundary conditions.
 \end{minipage}
\end{center}

\thispagestyle{empty}

\vfill
\newpage

\section{Introduction}
\label{sec:introduction}

When studying the Casimir effect \cite{casimir1948attraction} at
finite temperature \cite{fierz_attraction_1960,Mehra:1967wf}, one
cannot fail to be intrigued by the temperature inversion symmetry of
the partition function, originally derived using an image-source
construction of the Green's function \cite{Brown:1969na}. This result
has been re-discussed from various points of view including a
derivation in terms of Euclidean path integrals, Jacobi integrals and
Epstein zeta functions, while the Casimir energy has been related to
the thermodynamic potentials of the system
\cite{Dowker_1976,BALIAN1978165,10.1143/PTP.75.262,Ambjorn:1981xw,%
  Ambjorn:1981xv,Plunien:1987fr,Lutken:1988ge,Ford:1988gt,%
  PhysRevD.40.4191,Oikonomou:2006hm,Oikonomou:2007ww}. 

A natural question is then whether one may include another observable
in the partition function so as to enhance the temperature inversion
symmetry to transformations under the full modular group and complete
the parallel to the massless free boson on the torus with momentum
operator included. Whereas for two-dimensional conformal field
theories on the torus, the central charge has been related to the
Casimir energy of a field theory with boundary conditions in one
spatial dimension
\cite{Belavin:1984vu,Cardy:1986ie,Bloete:1986qm,Affleck:1986bv}, we
show here how techniques developed in that context can be applied in
the original setting of the Casmir effect in order to produce exact
results.

In order to address this question, we construct an exact equivalence, at the
level of finite temperature partition functions, between Maxwell's theory with
perfectly conducting parallel plate boundary conditions on $I_d \times \mathbb
R^2$ (or $I_d \times \mathbb T^2$), and a free massless scalar on
$S^1_{2d}\times \mathbb R^2$ (or $S^1_{2d} \times \mathbb T^2$). The exact
result for the extended partition function then readily follows from that of
this scalar field, which has been discussed in detail in \cite{Cappelli:1988vw},
and is inline with the analysis of modular invariance in
\cite{Shaghoulian:2015kta}.

That the spectrum of electromagnetism with perfectly conducting boundary
conditions corresponds to one scalar with Neumann and one scalar with Dirichlet
conditions, is discussed implicitly for instance in \cite{DeWitt:1975ys} section
2.4, and explicitly in \cite{Deutsch1979} (see also \cite{Plunien:1986ca},
section 3.2.2). This implies that, when taking perfectly conducting boundary
conditions into account, the computation cannot simply be done in terms of 2
polarizations with periodic or Dirichlet conditions as in empty space, but one
has to base it on $E$ and $H$ modes. It is the $E$ modes that contain the
additional $n_3=0$ modes because they satisfy Neumann conditions and have thus a
cosine expansion in the case of parallel plates. The scalar field theory with
periodic boundary conditions on the interval of double the length separating the
plates is then constructed out of these $E$ and $H$ modes.

Identifying the correct observable is straightforward in the scalar field
formulation. The corresponding expression in terms of electromagnetic fields is
somewhat harder to guess directly. The main point here is that the explicit map
between the scalar field and the original electromagnetic field is non-local in
space. As a consequence, the momentum in this direction in terms of the
electromagnetic field, the space integral of $T^{03}$ which is the observable
discussed in almost all other investigation of the Casimir effect, does not
correspond to scalar field momentum in the $x^3$ direction under the map.
Rather, as a direct computation using the detailed mode expansion shows, it
corresponds to the $x^3$-component of a suitably modified version of spin
angular momentum of light. This observable is usually not considered in the
context of the Casimir effect and that is the reason modular invariance beyond
temperature inversion symmetry cannot be discussed either.

Our motivation for studying this question originated from an attempt to
understand the contribution of specific degrees of freedom related to
non-trivial boundary conditions to partition functions. More precisely, a pair
of perfectly conducting plates at constant $x^3$ requires Neumann conditions for
the third component of the electric field and the vector potential. This gives
rise to an additional polarization at zero value of the quantized transverse
momentum whose dynamics is governed by a free massless scalar field in 2+1
dimensions. Its contribution to the partition function scales with the area of
the plates and provides the leading correction at low temperature to the zero
temperature Casimir result for the free energy
\cite{Barnich:2019xhd,Barnich2019}. Just as the contribution of the mode at
$n=0$ in the Fourier expansions of a free boson on a torus is essential to
achieve modular invariance of the partition function
\cite{Polchinski:1985zf,ItzyksonZuber1986}, so is the contribution of this
lower-dimensional massless scalar field in the current context.

\section{Capacitor partition function}
\label{sec:parall-plate-capact}

The partition function for perfectly conducting parallel plates of
area $L^2$ separated by a distance $d$, such that $L\gg d$, is given
by \cite{fierz_attraction_1960,Mehra:1967wf,Brown:1969na} (see
e.g.~\cite{Plunien:1986ca,Bordag:2009zzd} for reviews)
\begin{equation}
  \label{eq:1}
  \ln Z(t)=\frac{L^2}{d^2}\Big[\frac{\pi^2
    t}{360}-2t f({1}/{t})+\frac{\pi^2}{360
    t^3}\Big],\quad t=\frac{\beta}{2d}, 
\end{equation}
and
\begin{equation}
  \label{eq:2}
  f({1}/{t})=-\frac{1}{4\pi^2}\sum^\infty_{l,m=1}
  \frac{1}{(t^2l^2+m^2)^2},
\end{equation}
or, equivalently, when using a Sommerfeld-Watson transformation,
\begin{equation}
    \label{eq:7}
    f({1}/{t})= -\sum_{l=1}^\infty \Big[\frac{1}{16\pi t^3l^3}
    \coth (\pi l
    t)+\frac{1}{16 t^2l^2 \sinh^2(\pi l t)}
    \Big]+\frac{\pi^2}{720 t^4}.  
\end{equation}
The next step \cite{Lutken:1988ge,PhysRevD.40.4191} is to extend the
sum over all integers $(l,m)$ except for $(0,0)$ and to recognize the
relevant Epstein zeta function,
\begin{equation}
  \label{eq:14}
  {\mathfrak Z}(2;t^2,1)=\sum_{(l,m)\in \mathbb Z^2/{(0,0)}}\frac{1}{(l^2t^2+m^2)^2}, 
\end{equation}
in order to write the result as 
\begin{equation}
  \label{eq:15}
  \ln {Z}(t)=\frac{L^2t}{d^28\pi^2}{\mathfrak Z}(2;t^2,1),
\end{equation}
or, in terms of the free energy, 
\begin{equation}
  \label{eq:9}
  {F}(t)=-\frac{L^2}{d^3 16\pi^2}{\mathfrak Z}(2;t^2,1).
\end{equation}
The expression for the partition function in \eqref{eq:1} is in line
with the discussion in \cite{Brown:1969na}, but differs from the
corresponding result for the free energy in
\cite{Plunien:1986ca,Bordag:2009zzd} by the last term in \eqref{eq:1},
which is absent in the latter references. The reason is that the
latter approach includes the subtraction of the full free energy of
the black body in empty space, whereas this subtraction is limited to
the (divergent) zero temperature part in the former approach. This can
be done because the thermal part of the free energy is convergent. The
normalization condition chosen here is the one where the standard
black body result is recovered at large plate separation (see also
e.g.~\cite{Svaiter1992,Jauregui2006} for related discussions).

As pointed out in \cite{PhysRevD.40.4191}, an advantage of the
expression as given in \eqref{eq:1} is that the inversion symmetry,
$f({1}/{t})=t^{-4}f(t)$, established in \cite{Brown:1969na},
extends to the Epstein zeta function,
${\mathfrak Z}(2;{1}/{t^2},1)=t^4{\mathfrak Z}(2;t^2,1)$, and thus
turns into a symmetry of the full partition function and the free
energy,
\begin{equation}
  \label{eq:3}
  \ln {Z}({1}/{t}) = t^2 \ln {Z}(t),\quad
  {F}({1}/{t})=t^4{F}(t). 
\end{equation}
Up to exponentially suppressed terms, the high temperature expansion
$t\ll 1$ is
\begin{equation}
  \label{eq:4}
  \ln {Z}(t)\approx\frac{L^2}{d^2}[\frac{\pi^2}{360
    t^3}+\frac{1}{8\pi}\zeta(3)], 
\end{equation}
where the leading piece is the black body contribution
$\frac{V\pi^2}{45}\beta^{-3}$ while the sub-leading
temperature-independent contribution scales like the area; the low
temperature expansion $t \gg 1$ in turn is given by
\begin{equation}
  \label{eq:5}
  \ln {Z}(t)\approx \frac{L^2}{d^2}[\frac{\pi^2 t}{360}+\frac{1}{8\pi
    t^2}\zeta(3)], 
\end{equation}
where the second term is due to the lower-dimensional scalar as
described above.

\section{Boundary conditions: E and H modes}
\label{sec:e-h-modes}

Let us now provide some details on electromagnetism with Casimir
boundary conditions needed for our purpose. We work in radiation gauge
$A_0=0=\vec \nabla\cdot\vec A$ and implement from the outset the
constraints $\pi^0=0=\vec\nabla \cdot\vec \pi$. Perfectly conducting
large parallel plates of sides $L$ at $x^3=0$ and $x^3=d$ with unit
normal $\vec n$ require $\vec E\times \vec n=0=\vec B\cdot\vec n$ on
the plates, and periodic boundary conditions in the $x^a$ directions.
Let $i=1,2,3$, $a=1,2$,
\begin{equation}
  \label{eq:50}
  \begin{split}
    k_a=\frac{2\pi}{L}n_a, n_a\in \mathbb Z,\ k_3=\frac{\pi}{d}n_3,
    n_3\in \mathbb N,\ k=\sqrt{k_i k^i},\
    k_\perp=\sqrt{k_ak^a}, V=dL^2,\\
    \psi^H_{k}=\sqrt{\frac{2}{V}}e^{ik_a x^a}\sin k_3 x^3,\
    \psi^E_{k_a,0}=\frac{1}{\sqrt{V}}e^{ik_a x^a},\
    \psi^E_{k}=\sqrt{\frac{2}{V}}e^{ik_a x^a}\cos k_3 x^3,\ n_3\neq
    0.
  \end{split}
\end{equation}
Following for instance \cite{doi:10.1002/andp.19394270408} (see also
e.g.~\cite{DeWitt:1975ys,Deutsch1979} for closely related
discussions), one introduces 
\begin{equation}
  \label{eq:13}
   \begin{split}
  \phi^E=\sum_{n_i}\frac{1}{\sqrt{2 k} k k_\perp}
     [a^E_k\psi^E_k+{\rm c.c.}],\quad
  \pi^E=-i \sum_{n_i}\frac{1}{\sqrt{2 k} k_\perp}
     [a^E_k\psi^E_k-{\rm c.c.}],\\
  \phi^H=\sum_{n_i}\frac{1}{\sqrt{2 k} k_\perp}
     [a^H_k\psi^H_k+{\rm c.c.}],\quad
  \pi^H=-i \sum_{n_i}\frac{\sqrt k}{\sqrt{2} k_\perp}
     [a^H_k\psi^H_k-{\rm c.c.}],
   \end{split}
 \end{equation}
satisfying Neumann, respectively Dirichlet, conditions as well as the
Helmholtz equations
\begin{equation}
  \label{eq:20}
  (\Delta +k^2) \phi^\lambda=0=(\Delta +k^2)\pi^\lambda,
\end{equation}
with $\lambda=(E,H)$.  In these terms, the mode expansion of the
canonical pair $(\vec A,\vec \pi)$ and the associated electric and
magnetic fields $\vec E=-\vec \pi$, $\vec B=\vec \nabla\times \vec A$
is given by the sum of $E$ or transverse magnetic modes,
\begin{equation}
  \label{eq:10}
  \begin{array}{l}
    A^E_a=\partial_a\partial_3 \phi^E,\quad 
    A^E_3=(-\Delta+\partial_3^2)\phi^E,\\
    \pi^E_a=\partial_a\partial_3 \pi^E,\quad 
    \pi^E_3=(-\Delta+\partial_3^2)\pi^E,\\
    B^E_a=\epsilon_{ab}\partial^b(-\Delta)\phi^E,\quad 
    B^E_3=0,
  \end{array}
\end{equation}
where $\epsilon_{ab}$ is skew-symmetric with $\epsilon_{12}=1$,
and $H$ or transverse electric modes,
\begin{equation}
  \label{eq:10bis}
  \begin{array}{l}
    A^H_a=\epsilon_{ab}\partial^b\phi^H,\quad 
    A^H_3=0,\\
    \pi^H_a=\epsilon_{ab}\partial^b \pi^H,\quad 
    \pi^H_3=0,\\
    B^H_a=\partial_a\partial_3 \phi^H,\quad
    B^H_3=(-\Delta+\partial_3^2)\phi^H,
  \end{array}
\end{equation}
with the understanding that $a^H_{k_a,0}=0$.
In these variables, the first order electromagnetic action
\begin{equation}
  \label{eq:27}
  \begin{split}
    S=\int \D x^0\big[\int_V \D^3
    x\ \partial_0 {\vec A}\cdot \vec
  \pi-H\big],\quad H=\int_V \D^3x\ \frac{1}{2}(\vec E\cdot \vec
    E+\vec B\cdot \vec B),
  \end{split}
\end{equation}
is given by
\begin{equation}
  \label{eq:12}
  S=\int \D x^0 \sum_{n_i,\lambda}\big[\frac{1}{2i}(\partial_0
  a^{*\lambda}_ka^{\lambda}_k-a^{*\lambda}_k\partial_0 a^{\lambda}_k)-k
  a^{*\lambda}_ka^\lambda_k\big], 
\end{equation}
with $\lambda=(E,H)$. In particular, Poisson brackets are read off from
the kinetic term and oscillators do have the usual time
dependence. 

Action \eqref{eq:12} coincides with the mode expansion of an action
for two massless scalar fields, one with Neumann and one with
Dirichlet conditions. For later purposes, it is useful to introduce an
equivalent formulation in terms of a single free massless scalar
field, which satisfies periodic boundary conditions in $x,y$ in
intervals of length $L$ and in $x^3$ of length $2d$,
\begin{equation}
  \label{eq:25}
  \begin{split}
  \phi=\frac{1}{\sqrt{V_P}}\sum_{n_i}\frac{1}{\sqrt{2k}}[a_{k}e^{ik_j
    x^j}+{\rm c.c.}],\quad 
  \pi=\frac{-i}{\sqrt{V_P}}\sum_{n_i}\sqrt{\frac{k}{2}}[a_{k}e^{ik_j
    x^j}-{\rm c.c.}],
  \end{split}
\end{equation}
where $V_P=2dL^2$ with $n_3\in \mathbb Z$ as well, and 
\begin{equation}
  \label{eq:24a}
    a^{E}_{k_a,-k_3}=a^{E}_{k_a,k_3},\quad
    a^H_{k_a,-k_3}=-a^H_{k_a,k_3},\quad 
  a_{k}= \frac{a^E_{k}-ia^H_{k}}{\sqrt{2}},\ n_3
                   \neq 0,\quad
    a_{k_a,0}=  a^E_{k_a,0}. 
\end{equation}
With these definitions, the first order scalar field action
\begin{equation}
  \label{eq:28}
    S^S=\int \D x^0\big[\int_{V_P} \D^3x\big[\partial_0\phi\pi
    -H^S\big],\quad  
    H^S=\int_{V_P} \D^3x\ \frac{1}{2}(
    \pi^2+\vec\nabla\phi \cdot\vec\nabla \phi).
\end{equation}
agrees with the first order electromagnetic action \eqref{eq:27}
because its expression in terms of modes is given by the RHS of
\eqref{eq:12}.

\section{The observable}
\label{sec:observable}

In the equivalent scalar field formulation, we will show below that
the correct observable to be turned on in order to produce a real part
for the modular parameter $\tau$ and to consider full modular
transformations is linear momentum in the $x^3$ direction,
\begin{equation}
  \label{eq:30}
  P_3=-\int_{V_P} \D^3x\ \pi \partial_3\phi. 
\end{equation}
which, in terms of oscillators, is given by
\begin{equation}
  \label{eq:31}
  P_3=\sum_{n_i}k_3a^{*}_{k}a_{k}
  =\sum_{n_a,n_3>0}ik_3(a^{*H}_{k}a^E_{k}-a^{*E}_{k}a^H_{k}). 
\end{equation}
The action of this observable in electromagnetic terms can be inferred
from 
\begin{equation}
  \label{eq:6}
  \{\phi^E,P_3\}=(-\partial_3)(-\Delta)^{-\frac{1}{2}}\phi^H,\quad 
    \{\phi^H,P_3\}=(-\partial_3)(-\Delta)^{\frac{1}{2}}\phi^E,
\end{equation}
with similar relations holding for $\pi^\lambda$ by using
\eqref{eq:10} and \eqref{eq:10bis}. On the electromagnetic E and H
vector potentials, electric and magnetic fields, it acts like the curl
followed by an application of $(-\partial_3)(-\Delta)^{-\frac{1}{2}}$
and an exchange of $E$ and $H$:
\begin{equation}
  \label{eq:16}
  \begin{array}{l}
  \{\vec A^E,P_3\}=(-\partial_3)(-\Delta)^{-\frac{1}{2}}\vec B^H,\\
  \{\vec B^E,P_3\}=(-\partial_3)(-\Delta)^{-\frac{1}{2}}\partial_0\vec
  E^H,\\
  \{\vec E^E,P_3\}=(-\partial_3)(-\Delta)^{-\frac{1}{2}}(-\partial_0)\vec
    B^H,
  \end{array}
\end{equation}
with the transformations of the $H$ fields obtained by exchanging $E$
and $H$ in the above. Similarly, one may show by direct computation
that the observable is given by
\begin{equation}
  \label{eq:11}
  P_3=\int_V\D^3x\ \epsilon^{ab}(\sqrt{-\Delta}A^E_a\pi^H_b+
  \sqrt{-\Delta}A^H_a\pi^E_b). 
\end{equation}
Up to multiplication of each term in momentum space by $k$, this
observable is the component in the $x^3$ direction of
spin angular momentum, 
\begin{equation}
  \label{eq:17}
  \vec J=\int_V\D^3x\ \vec A\times \vec \pi,
\end{equation}
since one may show that
\begin{equation}
  \label{eq:8}
  J_3=\int_V \D^3x\
  \epsilon^{ab}(A^E_a\pi^H_b+A^H_a\pi^E_b). 
\end{equation}

\section{Extended partition function and modular properties}
\label{sec:zeta-funct-modul}

For the computation of the extended partition function, the fastet way
for our purpose is to follow \cite{Kapusta:1981aa} and to start from
the Hamiltonian path integral representation
\begin{equation}
  \label{eq:35}
  Z(\beta,\mu)={\rm Tr} e^{-\beta (\hat H-\mu \hat P_3)}=\int \prod d\phi \prod
  \frac{d\pi}{2\pi}\ e^{-S^E_H},
\end{equation}
where the sum is over periodic phase space path of period $\beta$, and
the Euclidean action is
\begin{equation}
  \label{eq:36}
  S^E_H=\int^\beta_0\D x^4\big[\int_{V_P}\D^3x (-i\partial_4\phi\pi) +(H^S-\mu
  P_3)\big]. 
\end{equation}
After integration over the momenta, this leads to
\begin{equation}
  \label{eq:37}
  Z(\beta,\mu)=\int \prod d\phi\  e^{-S^E_L}
\end{equation}
with
\begin{equation}
  \label{eq:38}
  S^E_L=\int^\beta_0\D x^4 \int_{V_P}\D^3x\
  \frac{1}{2} [(\partial_4\phi+i\mu\partial_3
  \phi)^2+\partial_j\phi\partial^j\phi]. 
\end{equation}
Following \cite{Hawking:1976ja} (see
also e.g.~\cite{DeWitt:1975ys,Dowker:1975tf} for earlier connected
work and \cite{doi:10.1142/2065} for a review), the evaluation of this
path integral is done by zeta function techniques.

Except for the replacement
$\partial_4\to \partial_4+i\mu\partial_3$, the operator in the
action is the Laplacian in 4 dimensions with periodic boundary
conditions in all directions. Since the eigenfunctions are
$e^{i(k_j x^j+\frac{2\pi n_4}{\beta}x^4)}$, the eigenvalues are
\begin{equation}
  \label{eq:39}
\lambda_{n_A}=  (2\pi)^2[\sum_a(\frac{n_a}{L})^2+(\frac{n_3}{2d})^2
  -(i\frac{n_4}{\beta}-\mu\frac{n_3}{2d})^2], 
\end{equation}
where $A=1,\dots 4$, the zeta function of the capacitor is
\begin{equation}
\zeta_C(s)={\sum_{n_A}}^\prime\lambda_{n_A}^{-s}\label{eq:42}
\end{equation}
where the prime means that the term with $n_A=0$ is excluded. Note
that the scalar field discussed at the end of the introduction
corresponds to the modes for which $n_3=0$ with $n_1,n_2,n_4$ not all
zero.  

The next step is to take the limit of large plate size $L$, and hence
to turn the sum over the transverse directions turns into integrals.
As in two-dimensional conformal field theories on the torus and also
in the context of QCD (see
e.g.~\cite{ROBERGE1986734,Alford:1998sd,PhysRevD.75.025003}), one now
uses a purely imaginary chemical potential $\mu=i\nu$ with $\nu$
real. After introducing the complex parameter
\begin{equation}
  \label{eq:41}
  \tau=\frac{\nu\beta+i\beta}{2d},
\end{equation}
and doing the integral in polar coordinates, the zeta function becomes
\begin{equation}
  \label{eq:40}
  \zeta_{C}(s)=\frac{L^2\beta^{2s-2}}{(2\pi)^{2s-1}}\int_{0}^\infty \D
  y \sum_{n_3,n_4}\frac{y}{[y^2+|\tau n_3+n_4|^2]^{s}}.
\end{equation}
When $n_3=0=n_4$ the integral is regulated through an infrared cut-off
$\epsilon$,
$\int^\infty_\epsilon \D y
y^{-2s+1}=-\frac{\epsilon^{2-2s}}{2-2s}$. This expression together
with its derivative both vanish in the limit at $s=0$ in the limit
$\epsilon\to 0$ and can thus be discarded. After performing the
integral for the remaining terms, the result may be written in terms
of a real analytic Eisenstein series (see
e.g.~\cite{fleig_gustafsson_kleinschmidt_persson_2018} for a recent
review),
\begin{equation}
  \label{eq:44}
  f_s(\tau)=\sum_{(m,n)\in\mathbb Z^2/(0,0)}
  \frac{[\mathfrak{Im}(\tau)]^s}{|m\tau+n|^{2s}}
\end{equation}
as 
\begin{equation}
  \label{eq:43}
  \zeta_{C}(s)=-
  \frac{L^2[\mathfrak{Im}(\tau)]^{s-1}
    f_{s-1}(\tau)}{(2-2s)(2\pi)^{2s-1}(2d)^{2-2s}}. 
\end{equation}
Using the inversion formula,
\begin{equation}
  \label{eq:45}
  \pi^{-s}\Gamma(s)f_s(\tau)=\pi^{s-1}\Gamma(1-s)f_{1-s}(\tau),
\end{equation}
then yields
\begin{equation}
  \label{eq:46}
  \zeta_{C}(s)=-
  \frac{L^2\Gamma(2-s)[\mathfrak{Im}(\tau)]^{s-1}
    f_{2-s}(\tau)}{(1-s)2^{2s}\pi^2(2d)^{2-2s}\Gamma(s-1)}
\end{equation}
Since $\Gamma(s-1)^{-1}=-s+O(s^2)$, it follows that
$\zeta_{C}(0)=0$, in which case $\ln Z(\tau)=\frac{1}{2} \zeta'_C(0)$
is explicitly given by 
\begin{equation}
  \label{eq:47}
  \ln Z(\tau)=\frac{L^2}{8\pi^2d^2}\frac{f_2(\tau)}{\mathfrak{Im}(\tau)}. 
\end{equation}
The behaviour of the partition function under modular transformations
\begin{equation}
  \label{eq:18}
  \tau\rightarrow \tau'=\frac{a\tau+b}{c\tau+d},\quad ad-bc=1,
\end{equation}
where $a,b,c,d\in\mathbb{Z}$, then follow directly from modular
invariance of $f_2(\tau)$,
\begin{equation}
  \label{eq:19}
  \ln Z(\tau')=|c\tau+d|^2\ln{Z}(\tau).
\end{equation}
The previous result \eqref{eq:15} corresponds to vanishing chemical
potential $\nu=0=\mathfrak{Re}(\tau)$ in which case the temperature
inversion formula is recovered for $a=0=d$, $b=-c=1$.

\section{Discussion}
\label{sec:discussion}

The main result of the paper is the exact computation in equation
\eqref{eq:47} of the partition function for electromagnetism between
two perfectly conducting parallel plates and the operator $P_3$ of
\eqref{eq:31}, or equivalently \eqref{eq:11}, turned on, in terms of
the real analytic Eisenstein series $f_2(\tau)$. This result is the
analog of the computation of the partition function with momentum
operator included, of the free boson on the torus given by
\begin{equation}
  Z_2(\tau)=\frac{1}{\sqrt{2\mathfrak{Im}(\tau)} |\eta(\tau)|^2},\label{eq:21}
\end{equation}
which is exactly modular invariant.

More details on a direct derivation in the operator formalism and on
an underlying infinite-dimensional symmetry algebra will appear
elsewhere. Other formulations making gauge invariance manifest will
also be explored. At this stage let us just point out that, after
having identified the modular parameter, one may write in the standard
way the contribution to the partition function that corresponds to the
vacuum energy, i.e., the first term in the RHS of \eqref{eq:5},
\begin{equation}
  {\frac{L^2\pi^2\beta}{d^3720}}=\ln{(q\bar q)^{-{c}/{24}}}=
  \frac{\pi\mathfrak{Im}(\tau) c}{6},\quad q=e^{2\pi i\tau}, \label{eq:48}
  \end{equation}
provided that the central charge of the planar capacitor is taken as 
\begin{equation}
  \label{eq:49}
  c=\frac{L^2\pi}{d^260}. 
\end{equation}

\section*{Acknowledgements}
\label{sec:acknowledgements}

The authors are grateful to M.~Bonte, G.~Giribet, A.~Kleinschmidt and
P.~Niro for comments. This work is supported by the F.R.S.-FNRS
Belgium through conventions FRFC PDR T.1025.14 and IISN 4.4503.15.


\providecommand{\href}[2]{#2}\begingroup\raggedright\endgroup

\end{document}